\documentclass[review]{elsarticle}

\usepackage{lineno,hyperref}
\usepackage{subfigure}
\usepackage{longtable}
\usepackage{verbatim}
\usepackage{xcolor}
\modulolinenumbers[5]
\usepackage{graphicx}
\usepackage{caption}
\usepackage{amsmath}

\begin{document}
\begin{frontmatter}

\title{Polarization Rank: A Study on European News Consumption on Facebook}

\author{Ana Luc\'{i}a Schmidt}
\address{Dept. of Environmental Sciences, Informatics and Statistics\\ Ca' Foscari University of Venice, Venice, Italy}
\ead{analucia.schmidt@unive.it}

\author{Fabiana Zollo}
\address{Dept. of Environmental Sciences, Informatics and Statistics \&\\ Center for the Humanities and Social Change\\Ca' Foscari University of Venice, Venice, Italy}

\author{Antonio Scala}
\address{ISC-CNR, Rome, Italy}

\author{Walter Quattrociocchi}
\address{Dept. of Environmental Sciences, Informatics and Statistics\\ Ca' Foscari University of Venice, Venice, Italy}

\begin{abstract}
The advent of WWW changed the way we can produce and access information. 
Recent studies showed that users tend to select information that is consistent with their system of beliefs, forming polarized groups of like-minded people around shared narratives where dissenting information is ignored.
In this environment, users cooperate to frame and reinforce their shared narrative making any attempt at debunking inefficient. 
Such a configuration occurs even in the consumption of news online, and considering that 63\% of users access news directly form social media
, one hypothesis is that more polarization allows for further spreading of misinformation.
Along this path, we focus on the polarization of users around news outlets on Facebook in different European countries (Italy, France, Spain and Germany).
First, we compare the pages' posting behavior and the users' interacting patterns across countries and observe different posting, liking and commenting rates.
Second, we explore the tendency of users to interact with different pages (i.e., selective exposure) and the emergence of polarized communities generated around specific pages.
Then, we introduce a new metric -- i.e., polarization rank --  to measure polarization of communities for each country.
We find that Italy is the most polarized country, followed by France, Germany and lastly Spain. Finally, we present a variation of the Bounded Confidence Model 
to simulate the emergence of these communities by considering the users' engagement and trust on the news. Our findings suggest that trust in information broadcaster plays a pivotal role against polarization of users online.
\end{abstract}

\begin{keyword}
Facebook, News Consumption, Misinformation, Polarization, Social Media
\end{keyword}
\end{frontmatter}

\section{Introduction}
The advent of social media changed the way we get informed and shape our opinion.
In 2016, \emph{post-truth} was selected by the Oxford Dictionaries as the word of the year.
The definition reads ``relating to or denoting circumstances in which objective facts are less influential in shaping public opinion than appeals to emotion and personal belief"€™, that is, we select information and interpretations adhering to our system of beliefs (confirmation bias).

This phenomenon is not new, our cognitive abilities have always been limited, and social media and the consequent disintermediated access to an unprecedented amount of information solely exacerbated the process.
Recent studies on massive datasets (376 million users) \cite{schmidt2017anatomy} showed that major disintermediated access to information is creating segregation of users into communities where they share a specific worldview and ignore dissenting information.
Confirmation bias dominates news consumption and informational cascades foster the emergence of polarized groups around shared narratives \cite{quattrociocchi2016echo, del2016spreading, del2017mapping, quattrociocchi2017inside, zollo2017debunking}.

Important results (that served to inform the Global risk report of the World Economic Forum in 2016 and 2017) pointed out the pivotal role of confirmation bias --i.e., the attitude of acquiring information coherently with the individual system of belief-- in viral processes as well as in the collective framing of narratives.
In particular, one of these works \cite{zollo2017debunking}, showing the inefficacy of debunking, convinced the Washington Post to close its weekly column dedicated to debunking false rumors \cite{dewey_2015}.

The process of acceptance of a claim (whether documented or not) may be altered by normative social influence or by the coherence with the individual system of beliefs as well-documented in the literature on cognitive and social psychology of communication \cite{nowak1990private, moscovici1982coming}.
At the extreme of the spectrum, conspiracy theorists tend to explain significant social or political aspects as plots conceived by powerful individuals or organizations, and with the so-called \textit{urban legends}€ they share an important characteristic: the object of the narratives are inevitably threatening the established social order or well-being and are always an indicator of what communities and social groups deeply fear \cite{franks2013conspiracy}.
These phenomena are evidently of great interest and can be considered as a sort of ``thermometer" of social mood. 
Since these kinds of arguments can sometimes involve the rejection of science, alternative explanations are invoked to replace the scientific evidence.
For instance, people who reject the link between HIV and AIDS generally believe that AIDS was created by the U.S. Government to control the African American population.


In this paper we focus on the interplay between users and news outlet on Facebook by comparing four European countries: France, Germany, Italy and Spain.
First, we compare the pages' posting behavior and the users' interacting patterns across countries and observe different posting, liking and commenting rates.
Second, we explore the tendency of users to interact with a variety of pages (i.e., selective exposure) and the polarized communities of pages that emerge from the users' consumption habits.
Then, we introduce a new method to calculate the percentage of polarized users when more than two communities are involved and thus rank the four countries accordingly.
We find that Italy is the most polarized country, followed by France, Germany and lastly Spain.
Finally, we present a variation of the Bounded Confidence Model \citep{deffuant2000mixing} to simulate the emergence of these communities by considering the users' engagement and trust on the news.

\section{Materials and Methods}
\subsection{Ethics Statement}
The data collection process was carried out using the Facebook Graph API \cite{fb_graph_api}, which is publicly available.
The pages from which we downloaded data are public Facebook entities and can be accessed by anyone.
Users' content contributing to such pages is  public unless users' privacy settings specify otherwise, and in that case their activity is not available to us.
	
\subsection{Data Collection}
We generated a list of top news sources, in their official language, of France, Germany, Italy and Spain.
The list for each country was compiled considering the Reuters Digital News Reports \cite{newman2015reuters}\cite{newman2016reuters}\cite{newman2017reuters}.
We then obtained the official Facebook page of each news outlet and proceeded to download all the posts made from 1st January 2015 to 31st December 2016, as well as all likes and comments that have been made on those posts.
The exact breakdown of the data can be seen in Tab.~\ref{tab1:dataset}, while the complete set of downloaded pages is reported in Tab.~\ref{tab6:pages_list} in the Supporting Information.

\begin{table}[!htb]
	\centering	
	\begin{tabular}{|c|c|c|c|c| }\hline
	        				&\textbf{France}&\textbf{Germany}&\textbf{Italy}& \textbf{Spain}\\\hline\hline
		\textit{Pages} 		& $65$ 			& $49$			& $54$			& $57$			\\ 
		\textit{Posts} 		& $1,008,018$ 	& $749,805$		& $1,554,817$	& $1,372,805$	\\
		\textit{Likes} 		& $419,371,366$ & $183,599,003$	& $409,243,176$	& $333,698,985$	\\
		\textit{Likers}		& $21,647,888$	& $14,367,445$ 	& $14,012,658$	& $32,812,007$	\\
		\textit{Comments} 	& $47,225,675$ 	& $31,881,407$ 	& $51,515,121$	& $34,336,356$	\\
		\textit{Commenters} & $5,755,268$ 	& $5,338,195$	& $4,086,351$	& $6,494,725$	\\
		\textit{Users} 		& $22,560,889$ 	& $15,564,360$	& $14,587,622$	& $34,383,820$	\\
		\textit{Population} & $66M$ 	 	& $81M$			& $62M$			& $46M$			\\\hline
	\end{tabular}
	\caption{\textbf{Dataset Breakdown.}
	Population according to the Reuters Digital News Report (2017) \cite{newman2017reuters}.	
	Likers is the number of people that gave at least one like.
	Commenters is the number of people that gave at least one comment.
	Users is the number of people that gave at least a like or comment.}
	\label{tab1:dataset}
\end{table}

\subsection{Preliminaries and Definitions}
In this section we provide a brief description of the main concepts and tools used in the analysis.

\subsubsection{Projection of Bipartite Graphs}
A bipartite graph is a triple $\mathcal{G}=(A,B,E)$ where $A=\left\{ a_{i}\,|\,i=1\dots n_{A}\right\} $ and $B=\left\{ b_{j}\,|\,j=1\dots n_{B}\right\} $ are two disjoint sets of vertices, and $E\subseteq A\times B$ is the set of edges, i.e. edges that exist only between vertices of sets $A$ and $B$.
The bipartite graph $\mathcal{G}$ is described by the rectangular matrix $M$ where $M_{ij}=1$, if an edge exits between $a_{i}$ and $b_{j}$, and $M_{ij}=0$ otherwise.

We consider bipartite networks in which the two disjointed set of nodes are users and Facebook pages.
That is $\mathcal{G}^{\kappa}=(P_{\kappa},U,E)$ where $P_{\kappa}$ is the set of Facebook pages of country $\kappa$ and $U$ is the set of users active on pages belonging to $P_{\kappa}$. 
Edges represent interactions among users and pages, that is, either likes or comments.

As an example, a like given to a post on page $p$ constitutes a link between the user $u$ and the page $p$ so $M_{p,u}=1$.
We can then build the co-occurrence matrices $C^{P_{\kappa}}=MM^{T}$ and $C^U=M^{T}M$ that quantify, respectively, the number of common neighbors between two vertices of $P_{\kappa}$ or $U$.

Only two graphs per country will be relevant for the analyses, $\mathcal{G}_L^{\kappa}$ and $\mathcal{G}_C^{\kappa}$.
These are the result from the projection $C^{P_{\kappa}}$ of two bipartite graphs: one considering the users' liking activity ($\mathcal{G}^{\kappa}_L$) and another considering the comments ($\mathcal{G}^{\kappa}_C$).

\subsubsection{Community Detection Algorithms}
Community detection algorithms serve to identify groups of nodes in a network.
In this work we apply three different community detection algorithms.

\begin{enumerate}
	\item \textit{FastGreedy} (FG).
	It takes an agglomerative bottom-up approach: initially each vertex belongs to a separate community and, at each iteration, the communities are merged in a way that yields the largest increase in the current value of modularity \cite{clauset2004finding}.
	The algorithm stops when it is no longer possible to further increase the modularity.
	Due to its speed and its lack of parameters in need of tuning, this algorithm will be the main reference to compare against the partitions resulting from the application of other community detection algorithms.
 
	\item \textit{Multilevel} (ML).
	It uses a multi-level optimization procedure for the modularity score \cite{blondel2008fast}.
	It takes a bottom-up approach where each vertex initially belongs to a separate community and in each step, unlike FastGreedy, vertices are reassigned in order to achieve the highest modularity.	
		
	\item \textit{Spinglass} (SG).
	It interprets the problem of community detection as one of finding the ground state of an infinite range spin-glass.
	In this algorithm, the community structure of the network would be the spin configuration that minimizes the energy of the spin glass, with the spin states being the community indices \cite{newman2004finding}\cite{reichardt2006statistical}.	
\end{enumerate}

To compare the various community partitions and the similarity between different clustering methods, we use the Rand index \cite{rand1971objective}, where a comparison between two partitions yields a value between 0 and
1, such that 0 indicates that there is no agreement on any vertex between the two partitions, whereas 1 indicates that the partitions are exactly the same.

\section{Results and Discussion}
\subsection{Attention Patterns}
As a first step we characterize how different countries consume news on Facebook.
We focus particularly on the allowed users' actions through the entire period of the data collection: \textit{likes}, \textit{shares} and \textit{comments}.
Naturally, each action has a prescribed meaning.
A \textit{like} represents a positive feedback to a post; a \textit{share} expresses the user's desire to increase the visibility of a given piece of information; and a \textit{comment} is the way in which online collective debates take form.
Therefore, comments may contain negative or positive feedback with respect to a post.

\begin{figure}[!htb]
	\centering
	\includegraphics[width=1\linewidth]{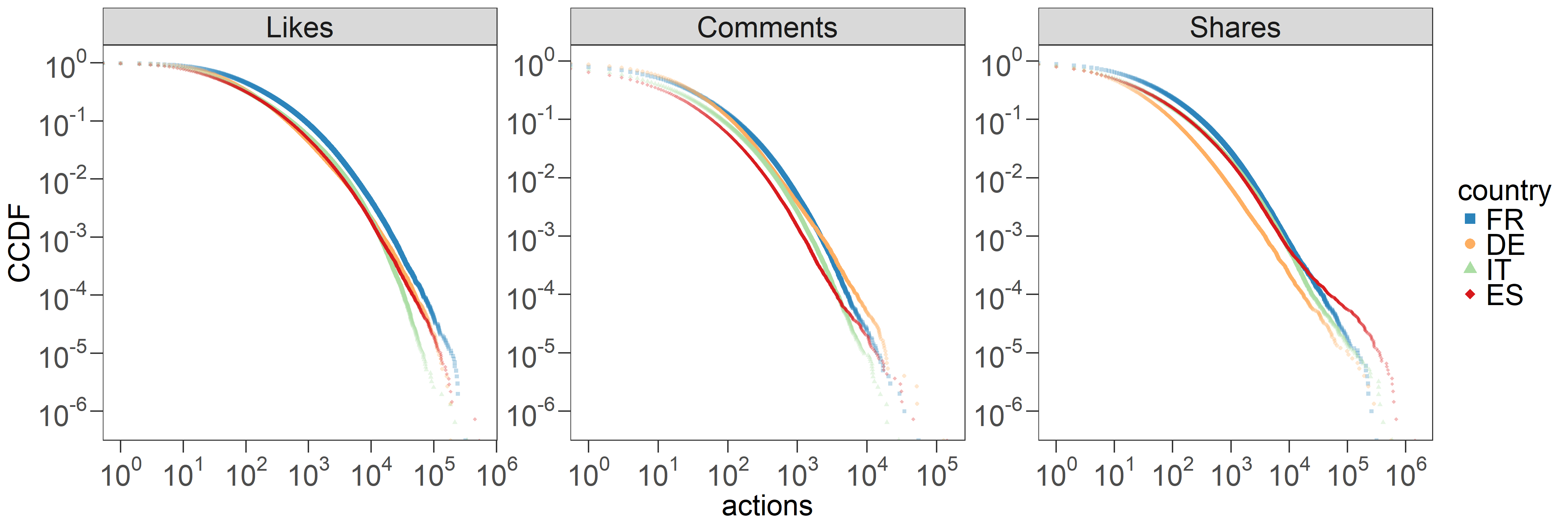}
	\caption{\textbf{ Complementary Cumulative Distribution Function of the comments, likes and shares received by the posts of each country.}}
	\label{fig1:ccdf_posts}
\end{figure}

In Fig.~\ref{fig1:ccdf_posts} we show the distribution of the number of likes, comments and shares received by the posts belonging to each country.
As seen from the plots, all the distributions are heavy-tailed, that is, they are best fitted by power laws (as shown in Tab.~\ref{tab2a:fit_posts}) and possess similar scaling parameters with some notable differences when looking at the number of comments and likes (Tab.~\ref{tab2b:powerlaw_posts}).

\begin{table}[!htb]
	\centering	
	\footnotesize
	\begin{tabular}{|c|c|c|c|c|c| }\hline
 		&\textbf{Action}&\textbf{Poisson}&\textbf{Log-Normal}&\textbf{Exponential}&\textbf{PowerLaw}\\\hline
		FR 		& comment 		& $-81,474,887$	& $-4,094,569$	& $-5,042,908$	& $-10,467$ \\ 
		DE 		& comment		& $-53,857,610$	& $-3,208,655$	& $-3,692,816$	& $-124,780$ \\
		IT 		& comment		& $-92,959,791$	& $-5,353,204$	& $-7,158,219$	& $-9,815$ \\
		ES 		& comment		& $-64,633,469$	& $-4,192,227$	& $-5,896,527$	& $-28,449$ \\ \hline
		FR 		& like 			& $-716,163,037$& $-6,463,931$	& $-7,135,298$	& $-9,793$ \\
		DE 		& like			& $-336,233,651$& $-4,429,366$	& $-4,906,736$	& $-243,938$ \\
		IT 		& like			& $-732,132,678$& $-9,034,577$	& $-10,276,500$	& $-12,514$ \\
		ES 		& like			& $-625,371,478$& $-7,905,112$	& $-8,978,996$	& $-34,532$ \\ \hline
		FR 		& share 		& $-302,119,999$& $-5,029,592$	& $-6,102,954$	& $-68,981$ \\
		DE 		& share			& $-100,787,846$& $-2,972,740$	& $-3,809,317$	& $-37,466$ \\
		IT 		& share			& $-399,573,409$& $-6,760,982$	& $-8,902,324$	& $-24,265$ \\
		ES 		& share			& $-456,628,686$& $-5,852,126$	& $-7,960,407$	& $-128,667$ \\ \hline
	\end{tabular}
	\caption{\textbf{Maximum-Likelihood fit of the actions received by the posts of each country.}
	FR: France, DE: Germany, IT: Italy, ES: Spain.}	
	\label{tab2a:fit_posts}
\end{table}

\begin{table}[!htb]
	\centering	
	\begin{tabular}{|c|c|c|c|c|c|c| }\hline
&\multicolumn{2}{|c|}{\textbf{Comments}}&\multicolumn{2}{|c|}{\textbf{Likes}}&\multicolumn{2}{|c|}{\textbf{Shares}} \\ \hline
	&$\hat{X}_{min}$&$\hat{\alpha}$&$\hat{X}_{min}$&$\hat{\alpha}$&$\hat{X}_{min}$&$\hat{\alpha}$\\ \hline
\textbf{FR}		& $1,929$	& $3.44$	& $23,338$	& $3.09$	& $2,498$	& $2.63$ \\ 
\textbf{DE}		& $315$		& $2.63$	& $1,132$	& $2.25$	& $1,084$	& $2.45$ \\ 
\textbf{IT}		& $1,736$	& $3.63$	& $15,519$	& $3.71$	& $5,753$	& $2.79$ \\ 
\textbf{ES}		& $733$		& $3.10$	& $8,491$	& $2.89$	& $1,508$	& $2.47$ \\ \hline
	\end{tabular}
	\caption{\textbf{Powerlaw fit of the actions received by the posts of each country.}}	
	\label{tab2b:powerlaw_posts}
\end{table}

\begin{figure}[!htb]
	\centering
	\includegraphics[width=1\linewidth]{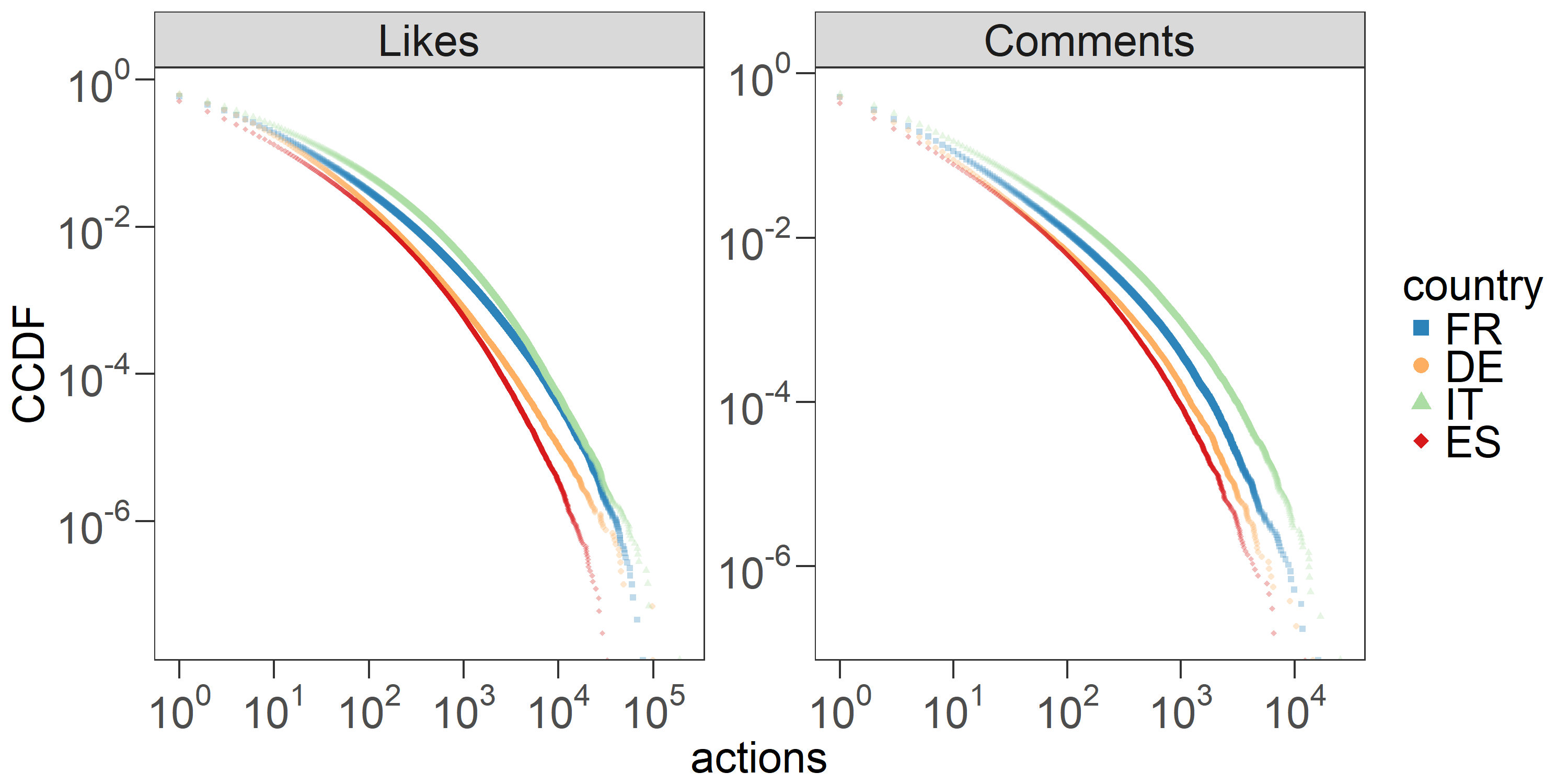}
	\caption{\textbf{Complementary Cumulative Distribution Function of the users' likes and comments of each country.}}
	\label{fig2:ccdf_users}
\end{figure}

We continue our analysis by examining how users from each country interact with the pages.
In Fig.~\ref{fig2:ccdf_users}, we show the distribution of the number of likes and comments given by the users according to each country.
Once again, all the distributions are heavy-tailed, as seen in Tab.~\ref{tab3a:fit_users}, with some notable differences in their scaling parameters when considering the commenting activity of the users of the different countries (Tab.~\ref{tab3b:powerlaw_users}).

\begin{table}[!htb]
	\centering	
	\footnotesize
	\begin{tabular}{|c|c|c|c|c|c| }\hline
&\textbf{Action}&\textbf{Poisson}&\textbf{Log-Normal}&\textbf{Exponential}&\textbf{PowerLaw}\\\hline
\textbf{FR} 	& comment 		& $-96,157,561$		& $-13,916,586$	& $-17,869,083$	& $-2,433$ \\ 
\textbf{DE} 	& comment		& $-57,870,795$		& $-11,951,470$	& $-14,878,272$	& $-1,268,430$ \\
\textbf{IT} 	& comment		& $-114,865,937$	& $-10,972,733$	& $-14,442,076$	& $-81,786$ \\
\textbf{ES} 	& comment		& $-62,141,913$		& $-13,638,119$	& $-17,309,835$	& $-11,920,701$ \\ \hline
\textbf{FR} 	& like 			& $-1,042,576,644$	& $-63,945,214$	& $-85,808,958$	& $-643,618$ \\
\textbf{DE} 	& like			& $-377,979,910$	& $-40,766,648$	& $-50,972,666$	& $-13,788$ \\
\textbf{IT} 	& like			& $-985,441,955$	& $-45,609,241$	& $-61,296,249$	& $-26,385$ \\
\textbf{ES} 	& like			& $-720,112,905$	& $-83,156,334$	& $-108,917,647$& $-48,326$ \\ \hline
	\end{tabular}
	\caption{\textbf{Maximum-Likelihood fit of the users' different actions by country.}
	FR: France, DE: Germany, IT: Italy, ES: Spain.}	
	\label{tab3a:fit_users}
\end{table}

\begin{table}[!htb]
	\centering	
	\begin{tabular}{|c|c|c|c|c| }\hline
			&\multicolumn{2}{|c|}{\textbf{Comments}}&\multicolumn{2}{|c|}{\textbf{Likes}} \\ \hline
&$\hat{X}_{min}$	&$\hat{\alpha}$	&$\hat{X}_{min}$	&$\hat{\alpha}$ \\ \hline
	FR 			& $2,378$		& $4.07$		& $648$			& $2.45$ \\ 
	DE 			& $18$			& $2.17$		& $3,156$		& $3.02$ \\ 
	IT 			& $529$			& $2.70$		& $5,473$		& $3.26$ \\ 
	ES	 		& $1$			& $1.90$		& $1,876$		& $3.24$ \\ \hline
	\end{tabular}
	\caption{\textbf{Power law fit of users' attention patterns.}}	
	\label{tab3b:powerlaw_users}
\end{table}

\subsection{Selective Exposure}
The overall number of likes given by each user is a good proxy for their level of \textit{engagement} with the Facebook news pages.
The \textit{lifetime} of a user, meaning the period of time where the user started and stopped interacting with our set of pages, can be approximated by the time difference between the time-stamp of their latest and earliest liked post.
These measures could provide important insights about news consumption patterns, specifically, the variety of news sources consumed over time.

We say that a user has consumed a page in a given time window, if the user has at least one positive interaction with that page in that period, that is, the user liked a post made by that page.
We do not consider comments as a valid interaction for regular consumption because they have very diverse meanings and, dissimilar from the likes, they do not unambiguously represent positive feedback.
Thus, we can measure the collection of pages consumed in a weekly, monthly and quarterly basis while taking into account the activity (total number of likes) and lifetime time difference of their first and last liked post) of the users of each country.

\begin{figure}[!hbt]
	\includegraphics[width=1\linewidth]{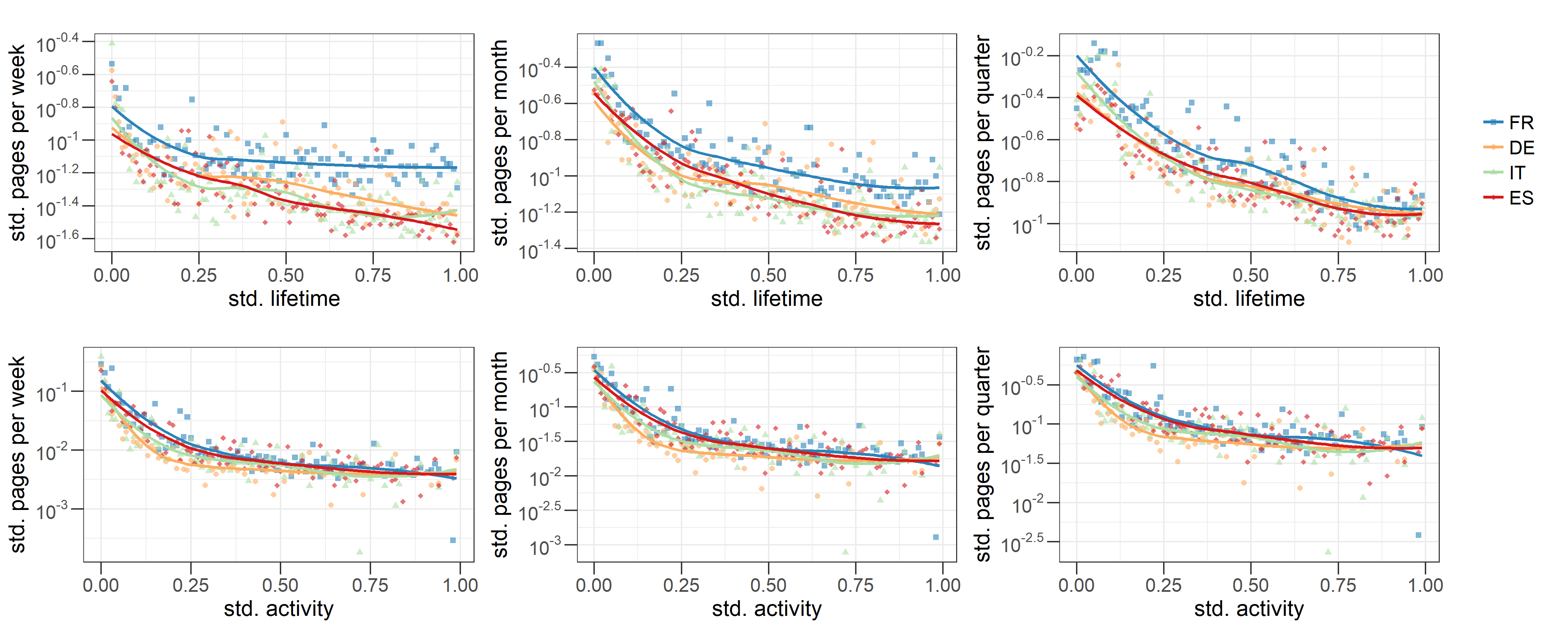}
	\caption{\textbf{Selective Exposure.}
	Maximum number of unique news sources that users with increasing levels of standardized lifetime (top)
	or standardized activity (bottom) interact with weekly, monthly and quarterly for each country.
	The user's lifetime corresponds to the normalized time difference between the time-stamp of their latest and
	earliest liked post. The user's activity corresponds to the number of likes given in their lifetime.}
	\label{fig3:selective_exposure}
\end{figure}

Fig.~\ref{fig3:selective_exposure} shows the number of news sources a user interacts with considering their lifetime and for increasing levels of engagement for each country. For a comparative analysis, we standardized between 0 and 1 the number of pages present in each country, as well as the lifetime and engagement over the entire user set. The results were calculated considering the quarterly (right), monthly (middle) and weekly (left) rates.

Note that, for all countries, users usually interact with a small number of news outlets and that higher levels of activity and longer lifetime correspond to a smaller variety of news sources being consumed.
We can also observe clear differences between the countries.
When considering the users' lifetime, France has clearly a more varied news consumption diet than the rest; and when considering the users' activity users in Germany consume consistently the less diverse set of news sources.
We can conclude that there is a natural tendency of the users to confine their activity to a limited set of pages, news consumption on Facebook is indeed dominated by selective exposure \cite{schmidt2017anatomy} and users from different countries display different rates for the decreasing variety of news outlets they consume.

\subsection{Emerging Communities}
User tendency to interact with few news sources might elicit page clusters.
To test this hypothesis, we first characterize the emergent community structure of pages according to the users' activity for each country $\kappa$ with $\kappa = \lbrace FR, DE, IT, ES \rbrace$. 
We project the users' page likes to derive the weighted graph $\mathcal{G}_L^\kappa$ (and $\mathcal{G}_C^\kappa$) in which nodes are pages and two pages are connected if a user likes (or comments on) both of them.
The weight of a link on a projected graph is determined by the number of users the two pages have in common.

\begin{table}[!h]
	\centering	
	\begin{tabular}{|c|c|c|c|c|c| }\hline
\textbf{$\mathcal{G}$} & $\kappa$ - \textbf{Country} &\textbf{Type} &\textbf{ML}	&\textbf{SG} \\ \hline\hline
\textbf{$\mathcal{G}_L^{FR}$} & France  & Likes		  				& $0.795$ 		& $0.796$\\
\textbf{$\mathcal{G}_L^{DE}$} & Germany & Likes    	  				& $0.771$		& $0.838$\\
\textbf{$\mathcal{G}_L^{IT}$} & Italy   & Likes    	 				& $0.982$		& $0.851$\\
\textbf{$\mathcal{G}_L^{ES}$} & Spain   & Likes    	  				& $0.923$		& $0.981$\\
\textbf{$\mathcal{G}_C^{FR}$} & France  & Comments 	  				& $0.918$		& $0.969$\\
\textbf{$\mathcal{G}_C^{DE}$} & Germany & Comments 	  				& $0.836$		& $0.925$\\
\textbf{$\mathcal{G}_C^{IT}$} & Italy   & Comments 	  				& $0.871$		& $0.903$\\
\textbf{$\mathcal{G}_C^{ES}$} & Spain   & Comments    				& $0.828$		& $0.817$\\\hline
	\end{tabular}
	\caption{\textbf{Algorithm comparison.}
	Comparison between the FastGreedy (FG) communities against the MultiLevel (ML) and SpinGlass (SG) communities for both likes and comments projections for every country.}	
	\label{tab4:fastgreedy_comparison}
\end{table}

We then apply the FastGreedy community detection algorithm to see if there are well-defined communities for each case.
To validate the community partitioning, we then compare the membership of other community detection algorithms using the Rand method \cite{rand1971objective} and find high level of similarity for all four countries (see Tab.~\ref{tab4:fastgreedy_comparison}).

We also compared the communities of $\mathcal{G}_L^\kappa$ and $\mathcal{G}_C^\kappa$ against each other using different community detection algorithms and find, overall, low levels of similarity (see Tab.~\ref{tab5:likes_vs_comments}).
This indicates that, for all four countries, the set pages users generally approve of (like), differ from the set of pages where they debate (comment).

\begin{table}[!h]
	\centering	
	\begin{tabular}{|c|c|c|c|c|c| }\hline
\textbf{Comparing} 			&$\kappa$ - \textbf{Country}	&\textbf{FG}&\textbf{ML}&\textbf{SG} \\ \hline\hline
\textbf{$\mathcal{G}_L^{FR}$-$\mathcal{G}_C^{FR}$} & France	& $0.514$	& $0.522$	& $0.545$\\
\textbf{$\mathcal{G}_L^{DE}$-$\mathcal{G}_C^{DE}$} & Germany& $0.528$	& $0.537$	& $0.518$\\
\textbf{$\mathcal{G}_L^{IT}$-$\mathcal{G}_C^{IT}$} & Italy  & $0.562$	& $0.560$	& $0.619$\\
\textbf{$\mathcal{G}_L^{ES}$-$\mathcal{G}_C^{ES}$} & Spain  & $0.555$	& $0.554$	& $0.625$\\\hline
	\end{tabular}
	\caption{\textbf{Likes and comments projections comparison.} Comparison of the communities detected in $\mathcal{G}_L^{\kappa}$ and $\mathcal{G}_C^{\kappa}$ of each country with FastGreedy (FG), MultiLevel (ML) and  SpinGlass (SG).}
	\label{tab5:likes_vs_comments}
\end{table}

\subsection{User Polarization}
By examining the activity of users across the various clusters  and measuring how they span across news outlets, we find that most users remain confined within specific groups of pages.
To understand the relationship between page groupings and user behavior, we measure the polarization of users with respect to the communities found for each country $\kappa$ where $\kappa=\lbrace FR, DE, IT, ES\rbrace$.

For a user with $K$ likes with $\sum_{i} k_{i}=K$ such that each $k_{i}$ belongs to the $i^{th}$ community ($i=1\ldots N$, where $N$ equals the number of communities).
The probability $\phi_{i}$ that the user belongs to the $i$-th community will then be $\phi_i = k_i/K$.
We can define the localization order parameter $L$ as:

\begin{equation}
L\left[\phi\right]=\frac{\left({\displaystyle \sum_{i}}\phi_i^{2}\right)^{2}}{{\displaystyle \sum_{i}\phi_i^{4}}}
\end{equation}

Thus, in the case in which the user only has likes in one community, $L=1$.
If a user, on the other hand, interacts equally with all the communities ($\phi_{i}=1/N$) then $L=N$; hence, $L$ \textit{counts} the communities.
Since we are considering many users, each with their likes $k_{i}$ and their frequency $\phi_{i}$, we can plot the probability distribution and the complementary cumulative distribution function of $L_\kappa$ along the user set of each country $\kappa$. 
This would allow for a fair comparison of the polarization of the users between countries.

\begin{figure}[!hbt]
	\centering
	\includegraphics[width=0.85\linewidth]{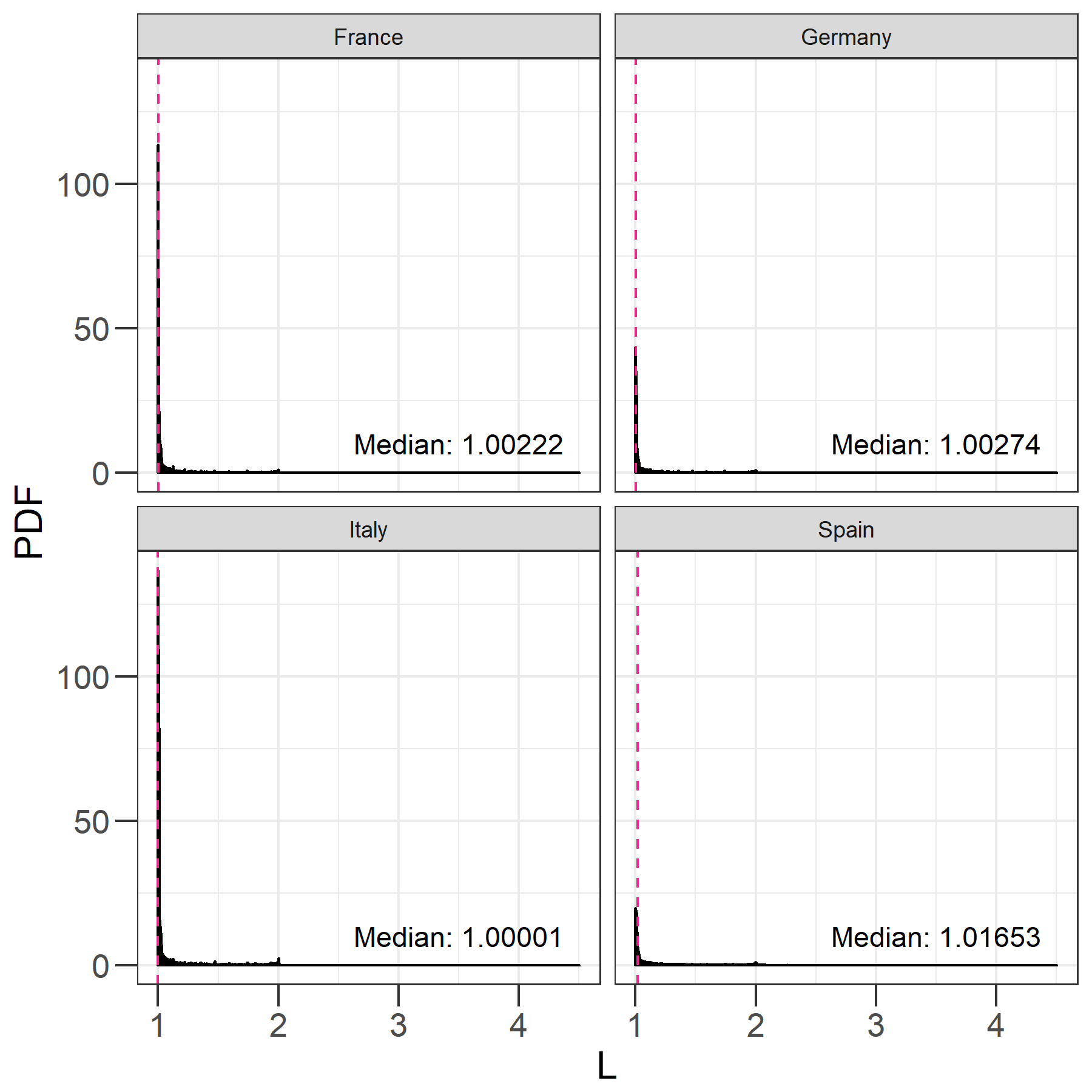}
	\caption{\textbf{Probability Density Function of $L$ for each country.} The dotted vertical line indicates the median value.}
	\label{fig4a:localization}
\end{figure}

\begin{figure}[!hbt]
	\centering
	\includegraphics[width=0.85\linewidth]{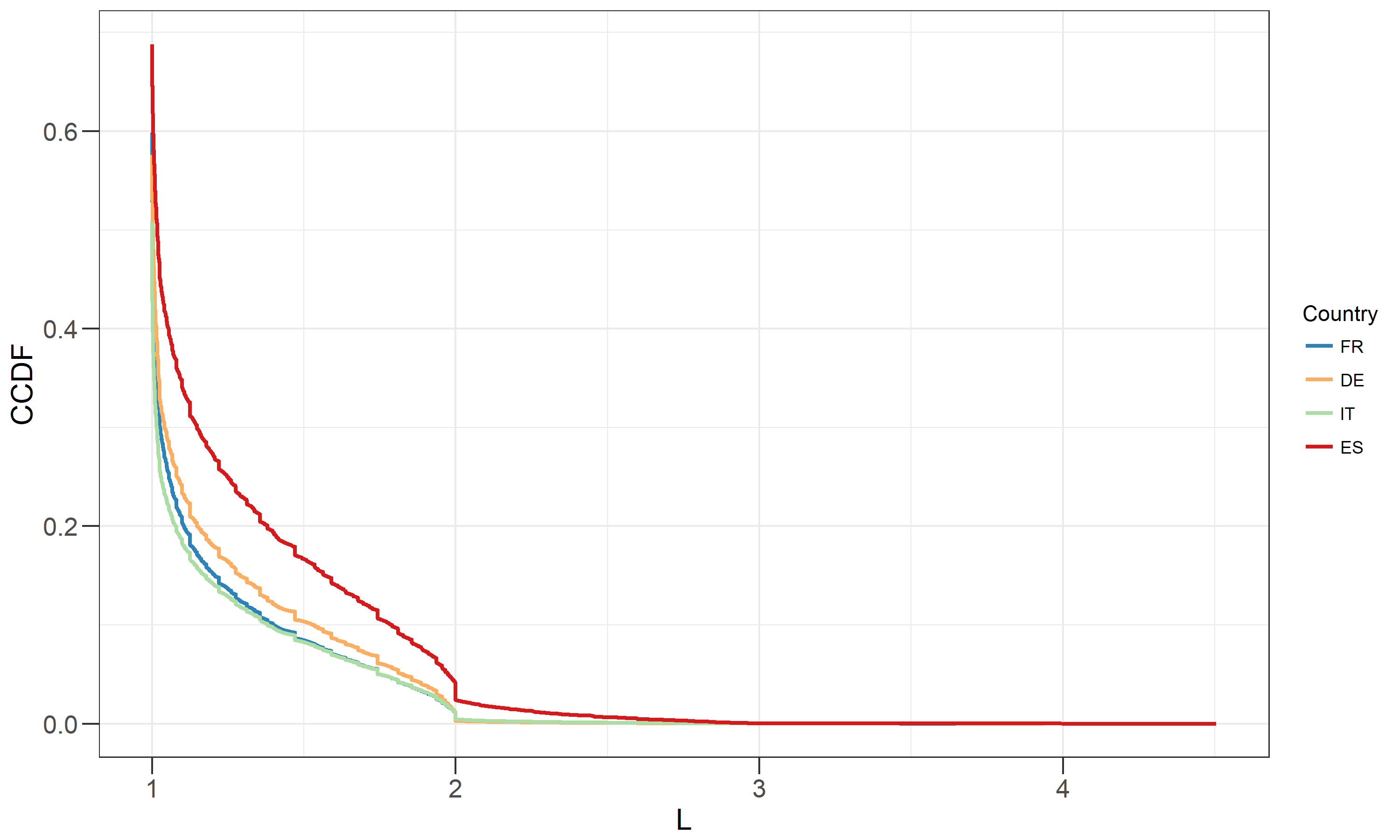}
	\caption{\textbf{Complementary Cumulative Distribution Function of $L$ for each country.}}
	\label{fig4b:ccdflocalization}
\end{figure}

For each country, Fig.~\ref{fig4b:ccdflocalization} shows the Complementary Cumulative Distribution Function of the localization $L$, and Fig.~\ref{fig4a:localization} shows the Probability Density Function.
Both figures consider only users with at least 10 likes.

As we can see in Fig.~\ref{fig4a:localization}, the densities are well behaved, that is, present a single peak around 1.
By looking at the CCDF of each country, we can rank the four countries from the one with least polarized users to the one with the most: Spain $(median=1.01653$), Germany ($median: 1.00274$), France ($median: 1.00222$) and Italy ($median: 1.00001$).

\subsection{The Model.}
In this section we provide a simple model of users' preferential attachment to specific sources that considers the users' \textit{trust} on the media as a parameter and reproduces the observed community structure.

The entities of our model are pages $p \in P$ and users $u \in U$.
Each page $p$ is characterized by a set of opinions (an editorial line) modelled as a real number $c_p$ that ranges $[0,1]$.
We assume that the $c_p$ values are uniformly distributed.
Each user $u$ has an initial opinion that is modelled as a real number $\theta_u$, which ranges between $[0,1]$ and it is uniformly distributed.
Each users $u$ also has a measure of trust in the media modelled by the real number $\tau_u$, which ranges between $[0,1]$.
User's trust will follow a truncated normal distribution.

We suppose $c_p$ and $\theta_u$ to be homogeneous such that the quantity $\vert c_p - \theta_u \vert$ is the distance between the opinion of user $u$ and the editorial line of page $p$.
We mimic confirmation bias by assuming that if user $u$ interacts with a page $p$ and the opinion distance $\vert c_p - \theta_u \vert$ is less than a given tolerance parameter $\Delta$, the preference of user $u$ will converge toward the editorial line of page $p$ according to the modified BCM \cite{deffuant2000mixing} equation:

\begin{equation}
\theta_u' = (1-\tau_u) \cdot \theta_u + \tau_u \cdot c_p
\end{equation}
\label{eq:BCM}

To mimic user activity we give each user $u$ an activity coefficient $a_u$ that represents the number of pages a user can visit.
Thus, the final opinion of a user will average the editorial lines of the pages the user likes.
If $\Omega$ is the set of $\left|\Omega\right|$ pages that matches the preferences of user $u$, then the average opinion will be:

\begin{align*}
\overline{\theta}_{u}	&=
\left( 1 -\tau_u \right) \overline{\theta}_{u} + \tau_u \left| \Omega \right|^{-1} \sum_{p\in\Omega}c_{p} \\
						&= \left| \Omega\right|^{-1} \sum_{p\in\Omega} c_{p} 
\end{align*}

To mimic the long tail distribution of our data we set the activity distribution to be power law distributed $p(a)\sim a^{-\gamma}$ with exponent $\gamma=3$.

We use numerical simulation to study our model.
A user randomly selects a subset of $P$ with which to interact.
The user likes a page only when $\vert c_p-\theta_u\vert <\Delta$.
When this occurs, the feedback mechanism reinforces the user's page preference using the trust parameter $\tau_u$ to control the extent of the feedback.
Thus the final opinion of a user will be the average of the editorial lines of the pages the user likes.

\begin{figure}[!htb]
	\centering
	\includegraphics[width=1\linewidth]{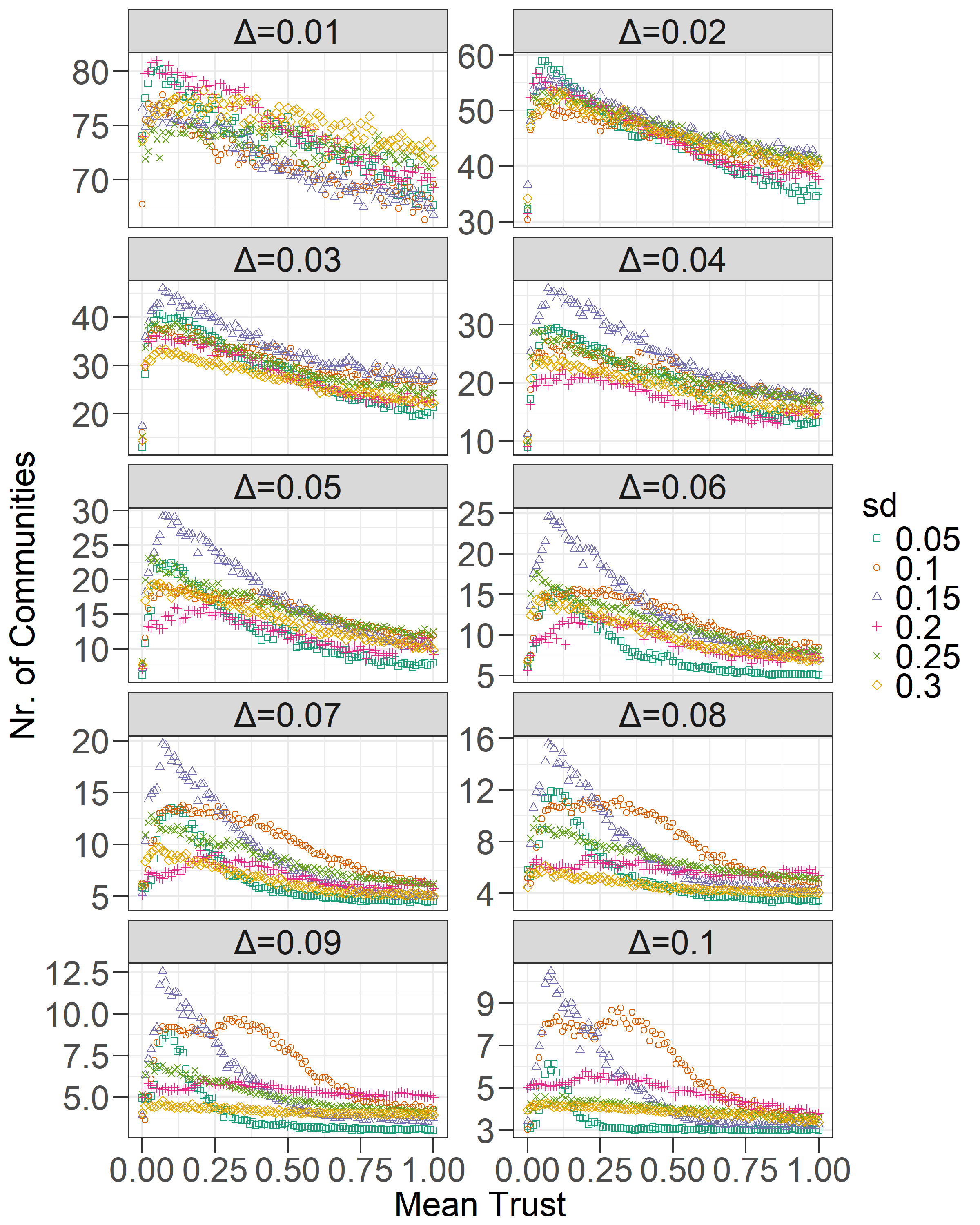}
	\caption{\textbf{Analysis of the synthetic pages-to-pages graph  $G_{\rm sim}^{p}$.}
	It shows the number of communities as a function of the mean user trust.}
	\label{fig5:model}
\end{figure}

When a user's opinion converges, we build in the bipartite graph $\mathcal{G}_{\rm sim}=(I,P,E_{\rm sim})$ where the set of edges $E_{\rm sim}$ are the couplings $(u,p)$ with which user $u$ likes page $p$.
Hence, $\mathcal{G}_{\rm sim}$ represents users interacting with their favorite pages, and from $\mathcal{G}_{\rm sim}$ we can build the projected graph $G_{\rm sim}^p$ that links the pages according their common users.

Figure \ref{fig5:model} shows an analysis of $G_{\rm sim}^{p}$ as a function of the mean values used for the truncated normal distribution that models the trust $\tau$, with different standard deviations and tolerance. Each point of the simulation is averaged over 100 iterations.

We can see that increasing the tolerance $\Delta$ leads to a reduction of the number of communities, that is, agreement is reached faster and polarization takes place.
Very low and very high values of user trust also display similar behavior.
Absolute trust or no trust in the media leads to fast polarization, either the user will trust what they read fully and change their opinion accordingly, or they won't.

The simulation displays an interesting behavior at $\tau=0.1$ where the number of communities formed by the users' consumption habits seem to peak.
This indicates that some skepticism might actually factor against polarization.
Users' who distrust the news they interact with, even when their opinions were similar, are more reluctant to further change their own beliefs.
Perhaps a solution for the issue of false and misleading narratives could be found by fostering critical readers.

\section{Discussion}
In this paper we use quantitative analysis to understand and compare the news consumption patterns of four European countries: France, Germany, Italy and Spain.
We show that while there are similarities in the consumption behaviours between the four countries, the posting and consumption behavior is not universal.

The results also show that all users, regardless of country, display selective exposure, that is, the more active a user is on Facebook the less variety of news sources they tend to consume.
This behavior is seen in all four countries, with different rates of selective exposure for each case.
News consumption on Facebook is dominated by selective exposure.

Additionally, we studied the cluster of news pages that emerge from the user's activity, and found that users, regardless of their nationality, are polarized.
We then measure the polarization of the users of each country, and ranked them accordingly, finding that Italy presents the most polarized users, followed by France, Germany and finally Spain.
Further studies might gain insights into the reasons behind the slight variations in consumption habits.

Finally, we introduce a variation on the Bonded Confidence Model \citep{deffuant2000mixing} that mimics the users' behavior of selective exposure taking into account user trust.
The simulation seems to indicate that users' who have some distrust of the news they interact with, even when the narrative presented conforms to their beliefs, are more reluctant to further change their own beliefs.
Thus, a tentative solution to mitigate user polarization might be found by fostering critical readers.

\section*{References}
\bibliography{mybib}

\newpage
\section*{Supporting Information}
\label{Supporting Information}
In this section we provide the list of all the downloaded pages.
Table \ref{tab6:pages_list} contains the 225 news pages that form the dataset.
Pages are identified by their name, website and Facebook ID, followed by the country code of their corresponding country.
The countries are indicated with their ISO Alpha-2 international code (FR: France, DE: Germany, IT: Italy, ES: Spain).

\renewcommand*{\arraystretch}{1.6}
\setlength{\LTleft}{-20cm plus -1fill}
\setlength{\LTright}{\LTleft}
\captionof{table}{List of pages of each country in the dataset.}
\label{tab6:pages_list}
\begin{footnotesize}
\begin{longtable}{ c | p{7cm} | l | c }
  \hline
  & Name and Website & Facebook ID & Community \\ 
  \hline
  1 & ARD - \scriptsize{\underline{\texttt{ard.de}}} & 48219766388 & DE \\ 
  2 & Augsburger Allgemeine Zeitung - \newline \scriptsize{\underline{\texttt{augsburger-allgemeine.de}}} & 121104385783 & DE \\ 
  3 & Badische Zeitung - \scriptsize{\underline{\texttt{badische-zeitung.de}}} & 177670301122 & DE \\ 
  4 & Berliner Morgenpost - \scriptsize{\underline{\texttt{morgenpost.de}}} & 46239931235 & DE \\ 
  5 & Berliner Zeitung - \scriptsize{\underline{\texttt{berliner-zeitung.de}}} & 137267732953826 & DE \\ 
  6 & Bild - \scriptsize{\underline{\texttt{bild.de}}} & 25604775729 & DE \\ 
  7 & B.Z. - \scriptsize{\underline{\texttt{bz-berlin.de}}} & 57187632436 & DE \\ 
  8 & Das Erste - \scriptsize{\underline{\texttt{daserste.de}}} & 176772398231 & DE \\ 
  9 & Der Spiegel - \scriptsize{\underline{\texttt{spiegel.de}}} & 38246844868 & DE \\ 
  10 & Der Tagesspiegel - \scriptsize{\underline{\texttt{tagesspiegel.de}}} & 59381221492 & DE \\ 
  11 & Der Westen - \scriptsize{\underline{\texttt{derwesten.de}}} & 243001859426137 & DE \\ 
  12 & Die Tageszeitung - \scriptsize{\underline{\texttt{taz.de}}} & 171844246207985 & DE \\ 
  13 & Die Welt - \scriptsize{\underline{\texttt{welt.de}}} & 97515118114 & DE \\ 
  14 & Die Zeit - \scriptsize{\underline{\texttt{zeit.de}}} & 37816894428 & DE \\ 
  15 & Express - \scriptsize{\underline{\texttt{express.de}}} & 172718036608 & DE \\ 
  16 & Focus - \scriptsize{\underline{\texttt{focus.de}}} & 37124189409 & DE \\ 
  17 & Frankfurter Allgemeine Zeitung - \scriptsize{\underline{\texttt{faz.net}}} & 346392590975 & DE \\ 
  18 & Frankfurter Rundschau - \scriptsize{\underline{\texttt{fr.de}}} & 134100583282150 & DE \\ 
  19 & Freie Presse - \scriptsize{\underline{\texttt{freiepresse.de}}} & 375109771472 & DE \\ 
  20 & Freitag - \scriptsize{\underline{\texttt{freitag.de}}} & 313744767921 & DE \\ 
  21 & GMX - \scriptsize{\underline{\texttt{gmx.net}}} & 187741777922914 & DE \\ 
  22 & Hamburger Abendblatt - \scriptsize{\underline{\texttt{abendblatt.de}}} & 121580125458 & DE \\ 
  23 & Hamburger Morgenpost - \scriptsize{\underline{\texttt{mopo.de}}} & 196072707519 & DE \\ 
  24 & Handelsblatt - \scriptsize{\underline{\texttt{handelsblatt.com}}} & 104709558232 & DE \\ 
  25 & Hannoversche Allgemeine Zeitung - \scriptsize{\underline{\texttt{haz.de}}} & 198530121257 & DE \\ 
  26 & Huffington Post DE - \scriptsize{\underline{\texttt{huffingtonpost.de}}} & 366193510165011 & DE \\ 
  27 & Junge Freiheit - \scriptsize{\underline{\texttt{jungefreiheit.de}}} & 13479664941 & DE \\ 
  28 & K\"{o}lner Stadt-Anzeiger - \scriptsize{\underline{\texttt{ksta.de}}} & 141063022950 & DE \\ 
  29 & Leipziger Volkszeitung - \scriptsize{\underline{\texttt{lvz.de}}} & 114360055263804 & DE \\ 
  30 & Mitteldeutsche Zeitung - \scriptsize{\underline{\texttt{mz-web.de}}} & 141558262607 & DE \\ 
  31 & n-tv online - \scriptsize{\underline{\texttt{n-tv.de}}} & 126049165307 & DE \\ 
  32 & Ostsee-Zeitung - \scriptsize{\underline{\texttt{ostsee-zeitung.de}}} & 374927701107 & DE \\ 
  33 & ProSieben Newstime - \scriptsize{\underline{\texttt{prosieben.de/tv/newstime}}} & 64694257920 & DE\\ 
  34 & Rheinische Post - \scriptsize{\underline{\texttt{rp-online.de}}} & 50327854366 & DE \\ 
  35 & RTL aktuell - \scriptsize{\underline{\texttt{rtluell.de}}} & 119845424729050 & DE \\ 
  36 & SAT1 Nachrichten - \scriptsize{\underline{\texttt{sat1.de/news}}} & 171663852895480 & DE \\ 
  37 & Schleswig-Holsteinischer Zeitungsverlag - \scriptsize{\underline{\texttt{shz.de}}} & 248528847673 & DE \\ 
  38 & Stern - \scriptsize{\underline{\texttt{stern.de}}} & 78766664651 & DE \\ 
  39 & Stuttgarter Nachrichten - \scriptsize{\underline{\texttt{stuttgarter-nachrichten.de}}} & 144537361776 & DE \\ 
  40 & Stuttgarter Zeitung - \scriptsize{\underline{\texttt{stuttgarter-zeitung.de}}} & 129349103260 & DE \\ 
  41 & S\"{u}ddeutsche Zeitung - \scriptsize{\underline{\texttt{sueddeutsche.de}}} & 215982125159841 & DE \\ 
  42 & tagesschau - \scriptsize{\underline{\texttt{tagesschau.de}}} & 193081554406 & DE \\ 
  43 & t-online - \scriptsize{\underline{\texttt{t-online.de}}} & 24897707939 & DE \\ 
  44 & WAZ - \scriptsize{\underline{\texttt{waz.de}}} & 117194401183 & DE \\ 
  45 & WEB.DE - \scriptsize{\underline{\texttt{web.de}}} & 56488242934 & DE \\ 
  46 & Wirtschafts Woche - \scriptsize{\underline{\texttt{wiwo.de}}} & 93810620818 & DE \\ 
  47 & Yahoo News DE - \scriptsize{\underline{\texttt{de.nachrichten.yahoo.com}}} & 166721106679241 & DE \\ 
  48 & ZDF - \scriptsize{\underline{\texttt{zdf.de}}} & 154149027994068 & DE \\ 
  49 & ZDF heute - \scriptsize{\underline{\texttt{heute.de}}} & 112784955679 & DE \\ 
  50 & 20 MINUTOS - \scriptsize{\underline{\texttt{20minutos.es}}} & 38352573027 & ES \\ 
  51 & ABC - \scriptsize{\underline{\texttt{abc.es}}} & 7377874895 & ES \\ 
  52 & Antena 3 - \scriptsize{\underline{\texttt{antena3.com}}} & 55353596297 & ES \\ 
  53 & Cadena Ser - \scriptsize{\underline{\texttt{cadenaser.com}}} & 15658775846 & ES \\ 
  54 & Canarias 7 - \scriptsize{\underline{\texttt{canarias7.es}}} & 85160277321 & ES \\ 
  55 & Cinco D\'{i}as - \scriptsize{\underline{\texttt{cincodias.elpais.com}}} & 36280712574 & ES \\ 
  56 & COPE - \scriptsize{\underline{\texttt{cope.es}}} & 15829535820 & ES \\ 
  57 & Cuatro news - \scriptsize{\underline{\texttt{cuatro.com/noticias}}} & 96876562265 & ES \\ 
  58 & Diario de C\'{a}diz - \scriptsize{\underline{\texttt{diariodecadiz.es}}} & 128335533904779 & ES \\ 
  59 & Diario de Ibiza - \scriptsize{\underline{\texttt{diariodeibiza.es}}} & 255177630236 & ES \\ 
  60 & Diario de Mallorca - \scriptsize{\underline{\texttt{diariodemallorca.es}}} & 155352736257 & ES \\ 
  61 & Diario de Navarra - \scriptsize{\underline{\texttt{diariodenavarra.es}}} & 103384039711468 & ES \\ 
  62 & El Comercio - \scriptsize{\underline{\texttt{elcomercio.es}}} & 64673887657 & ES \\ 
  63 & El Confidencial - \scriptsize{\underline{\texttt{elconfidencial.com}}} & 63830851925 & ES \\ 
  64 & El Confidencial Digital - \scriptsize{\underline{\texttt{elconfidencialdigital.com}}} & 202726949863885 & ES \\ 
  65 & El Correo - \scriptsize{\underline{\texttt{elcorreo.com}}} & 280982578099 & ES \\ 
  66 & El Correo Gallego - \scriptsize{\underline{\texttt{elcorreogallego.es}}} & 152802838075123 & ES \\ 
  67 & El D\'{i}a - \scriptsize{\underline{\texttt{eldia.es}}} & 165210860204301 & ES \\ 
  68 & ElDiario.es - \scriptsize{\underline{\texttt{eldiario.es}}} & 417471918268686 & ES \\ 
  69 & El Diario Monta\~{n}\'{e}s - \scriptsize{\underline{\texttt{eldiariomontanes.es}}} & 109434489075314 & ES \\ 
  70 & El Diario Vasco - \scriptsize{\underline{\texttt{diariovasco.com}}} & 91085818678 & ES \\ 
  71 & El Economista - \scriptsize{\underline{\texttt{eleconomista.es}}} & 56760767000 & ES \\ 
  72 & El Espa\~{n}ol - \scriptsize{\underline{\texttt{elespanol.com}}} & 693292367452833 & ES \\ 
  73 & El Mundo - \scriptsize{\underline{\texttt{elmundo.es}}} & 10407631866 & ES \\ 
  74 & El Norte de Castilla - \scriptsize{\underline{\texttt{elnortedecastilla.es}}} & 98474974005 & ES \\ 
  75 & El Pa\'{i}s - \scriptsize{\underline{\texttt{elpais.com}}} & 8585811569 & ES \\ 
  76 & El Peri\'{o}dico - \scriptsize{\underline{\texttt{elperiodico.com}}} & 93177351543 & ES \\ 
  77 & Expansi\'{o}n - \scriptsize{\underline{\texttt{expansion.com}}} & 93983931918 & ES \\ 
  78 & Faro de Vigo - \scriptsize{\underline{\texttt{farodevigo.es}}} & 123746764304270 & ES \\ 
  79 & Heraldo de Arag\'{o}n - \scriptsize{\underline{\texttt{heraldo.es}}} & 130012437016272 & ES \\ 
  80 & Hoy - \scriptsize{\underline{\texttt{hoy.es}}} & 85593393832 & ES \\ 
  81 & Ideal - \scriptsize{\underline{\texttt{ideal.es}}} & 64258697112 & ES \\ 
  82 & Informaci\'{o}n - \scriptsize{\underline{\texttt{diarioinformacion.com}}} & 410523955526 & ES \\ 
  83 & La Gaceta de Salamanca - \scriptsize{\underline{\texttt{lagacetadesalamanca.es}}} & 319669591452311 & ES \\ 
  84 & La Nueva Espa\~{n}a - \scriptsize{\underline{\texttt{lne.es}}} & 51837272861 & ES \\ 
  85 & La Opini\'{o}n de M\'{a}laga - \scriptsize{\underline{\texttt{laopiniondemalaga.es}}} & 80999977105 & ES \\ 
  86 & La Opini\'{o}n de Murcia - \scriptsize{\underline{\texttt{laopiniondemurcia.es}}} & 106647502704110 & ES \\ 
  87 & La Opini\'{o}n de Tenerife - \scriptsize{\underline{\texttt{laopinion.es}}} & 112238345503995 & ES \\ 
  88 & La Provincia - \scriptsize{\underline{\texttt{laprovincia.es}}} & 124641092828 & ES \\ 
  89 & La Raz\'{o}n - \scriptsize{\underline{\texttt{larazon.es}}} & 113080018770027 & ES \\ 
  90 & La Sexta - \scriptsize{\underline{\texttt{lasexta.com}}} & 39172614918 & ES \\ 
  91 & Las Provincias - \scriptsize{\underline{\texttt{lasprovincias.es}}} & 20810574989 & ES \\ 
  92 & La Vanguardia - \scriptsize{\underline{\texttt{lavanguardia.com}}} & 156552584408339 & ES \\ 
  93 & La Verdad - \scriptsize{\underline{\texttt{laverdad.es}}} & 120857625399 & ES \\ 
  94 & La Voz de Asturias - \scriptsize{\underline{\texttt{lavozdeasturias.es}}} & 101351926940208 & ES \\ 
  95 & La Voz De Galicia - \scriptsize{\underline{\texttt{lavozdegalicia.es}}} & 350393845757 & ES \\ 
  96 & Levante-EMV - \scriptsize{\underline{\texttt{levante-emv.com}}} & 106329485190 & ES \\ 
  97 & Libertad Digital - \scriptsize{\underline{\texttt{libertaddigital.com}}} & 141423087721 & ES \\ 
  98 & MSN Espa\~{n}a - \scriptsize{\underline{\texttt{msn.com/es-es}}} & 35966491049 & ES \\ 
  99 & Onda Cero - \scriptsize{\underline{\texttt{ondacero.es}}} & 99040469027 & ES \\ 
  100 & P\'{u}blico - \scriptsize{\underline{\texttt{publico.es}}} & 75084861845 & ES \\ 
  101 & QUE! - \scriptsize{\underline{\texttt{que.es}}} & 97090259641 & ES \\ 
  102 & RTVE - \scriptsize{\underline{\texttt{rtve.es}}} & 133623265400 & ES \\ 
  103 & Sur - \scriptsize{\underline{\texttt{diariosur.es}}} & 52107727250 & ES \\ 
  104 & Telecinco - \scriptsize{\underline{\texttt{telecinco.es}}} & 50353113909 & ES \\ 
  105 & \'{U}ltima Hora - \scriptsize{\underline{\texttt{ultimahora.es}}} & 114680095225282 & ES \\ 
  106 & Yahoo News ES - \scriptsize{\underline{\texttt{es.noticias.yahoo.com}}} & 284428852938 & ES \\ 
  107 & 20 Minutes - \scriptsize{\underline{\texttt{20minutes.fr}}} & 51555073310 & FR \\ 
  108 & Agence France-Presse - \scriptsize{\underline{\texttt{afp.com/fr}}} & 114100038626559 & FR \\ 
  109 & BFMTV - \scriptsize{\underline{\texttt{bfmtv.com}}} & 43896752783 & FR \\ 
  110 & Canal+ - \scriptsize{\underline{\texttt{canalplus.fr}}} & 144056732332683 & FR \\ 
  111 & Challenges - \scriptsize{\underline{\texttt{challenges.fr}}} & 79566127213 & FR \\ 
  112 & Charente Libre - \scriptsize{\underline{\texttt{charentelibre.fr}}} & 144375072241306 & FR \\ 
  113 & Charlie Hebdo - \scriptsize{\underline{\texttt{charliehebdo.fr}}} & 106626879360459 & FR \\ 
  114 & CNES Matin - \scriptsize{\underline{\texttt{cnewsmatin.fr}}} & 181111805243991 & FR \\ 
  115 & CNEWS - \scriptsize{\underline{\texttt{cnews.fr}}} & 76952916976 & FR \\ 
  116 & Corse Matin - \scriptsize{\underline{\texttt{corsematin.com}}} & 107249929306302 & FR \\ 
  117 & Courrier international - \scriptsize{\underline{\texttt{courrierinternational.com}}} & 142114104887 & FR \\ 
  118 & Dernieres Nouvelles d'Alsace - \scriptsize{\underline{\texttt{dna.fr}}} & 19004867327 & FR \\ 
  119 & FranceInfo - \scriptsize{\underline{\texttt{francetvinfo.fr}}} & 135112586936434 & FR \\ 
  120 & France Soir - \scriptsize{\underline{\texttt{francesoir.fr}}} & 53638966652 & FR \\ 
  121 & France T\'{e}l\'{e}visions - \scriptsize{\underline{\texttt{francetelevisions.fr}}} & 179086202130933 & FR \\ 
  122 & Huffington Post FR - \scriptsize{\underline{\texttt{huffingtonpost.fr}}} & 284129444969978 & FR \\ 
  123 & La Croix - \scriptsize{\underline{\texttt{la-croix.com}}} & 108828257010 & FR \\ 
  124 & La D\'{e}p\^{e}che du Midi - \scriptsize{\underline{\texttt{ladepeche.fr}}} & 271219815470 & FR \\ 
  125 & L'Alsace - Le Pays - \scriptsize{\underline{\texttt{lalsace.fr}}} & 181480351879611 & FR \\ 
  126 & La Montagne - \scriptsize{\underline{\texttt{lamontagne.fr}}} & 146949065315655 & FR \\ 
  127 & La Nouvelle R\'{e}publique du Centre Ouest - \newline \scriptsize{\underline{\texttt{lanouvellerepublique.fr}}} & 87693933163 & FR \\ 
  128 & La Provence - \scriptsize{\underline{\texttt{laprovence.com}}} & 119213845538 & FR \\ 
  129 & La R\'{e}publique des Pyrenn\'{e}es - \newline \scriptsize{\underline{\texttt{larepubliquedespyrenees.fr}}} & 148446219817 & FR \\ 
  130 & La R\'{e}publique du Centre - \scriptsize{\underline{\texttt{larep.fr}}} & 211082695569481 & FR \\ 
  131 & La Tribune - \scriptsize{\underline{\texttt{latribune.fr}}} & 18950434380 & FR \\ 
  132 & La Voix du Nord - \scriptsize{\underline{\texttt{lavoixdunord.fr}}} & 76635774021 & FR \\ 
  133 & Le Bien Public - \scriptsize{\underline{\texttt{bienpublic.com}}} & 106094599409 & FR \\ 
  134 & Le Courrier Picard - \scriptsize{\underline{\texttt{courrier-picard.fr}}} & 58080584133 & FR \\ 
  135 & Le Dauphin\'{e} Lib\'{e}r\'{e} - \scriptsize{\underline{\texttt{ledauphine.com}}} & 122601757780987 & FR \\ 
  136 & Le Figaro - \scriptsize{\underline{\texttt{lefigaro.fr}}} & 61261101338 & FR \\ 
  137 & Le Journal du Dimanche - \scriptsize{\underline{\texttt{lejdd.fr}}} & 246577183385 & FR \\ 
  138 & Le Monde - \scriptsize{\underline{\texttt{lemonde.fr}}} & 14892757589 & FR \\ 
  139 & Le Monde Diplomatique - \scriptsize{\underline{\texttt{monde-diplomatique.fr}}} & 34398236687 & FR \\ 
  140 & Le Nouvel Observateur - \scriptsize{\underline{\texttt{tempsreel.nouvelobs.com}}} & 198508090036 & FR \\ 
  141 & Le Parisien - \scriptsize{\underline{\texttt{leparisien.fr}}} & 36550584062 & FR \\ 
  142 & Le Point - \scriptsize{\underline{\texttt{lepoint.fr}}} & 49173930702 & FR \\ 
  143 & Le Populaire du Centre - \scriptsize{\underline{\texttt{lepopulaire.fr}}} & 240500052515 & FR \\ 
  144 & Le Progr\`{e}s - \scriptsize{\underline{\texttt{leprogres.fr}}} & 104985642868265 & FR \\ 
  145 & Le R\'{e}publicain Lorrain - \scriptsize{\underline{\texttt{republicain-lorrain.fr}}} & 142638581774 & FR \\ 
  146 & Les \'{E}chos - \scriptsize{\underline{\texttt{lesechos.fr}}} & 123440511000645 & FR \\ 
  147 & L'Est R\'{e}publicain - \scriptsize{\underline{\texttt{estrepublicain.fr}}} & 190366851765 & FR \\ 
  148 & Le T\'{e}l\'{e}gramme - \scriptsize{\underline{\texttt{letelegramme.fr}}} & 97539957978 & FR \\ 
  149 & L'Express - \scriptsize{\underline{\texttt{lexpress.fr}}} & 9359316996 & FR \\ 
  150 & L'Humanit\'{e} - \scriptsize{\underline{\texttt{humanite.fr}}} & 254585183694 & FR \\ 
  151 & Lib\'{e}ration - \scriptsize{\underline{\texttt{liberation.fr}}} & 147126052393 & FR \\ 
  152 & L'Ind\'{e}pendant - \scriptsize{\underline{\texttt{lindependant.fr}}} & 52697519148 & FR \\ 
  153 & L'internaute - \scriptsize{\underline{\texttt{linternaute.com}}} & 156569814356922 & FR \\ 
  154 & L'Opinion - \scriptsize{\underline{\texttt{lopinion.fr}}} & 445890365491209 & FR \\ 
  155 & L'Union - \scriptsize{\underline{\texttt{lunion.fr}}} & 100163350071823 & FR \\ 
  156 & Marianne - \scriptsize{\underline{\texttt{marianne.net}}} & 369717525444 & FR \\ 
  157 & Mediapart - \scriptsize{\underline{\texttt{mediapart.fr}}} & 116070051527 & FR \\ 
  158 & Metro France - \scriptsize{\underline{\texttt{lci.fr}}} & 411124728976705 & FR \\ 
  159 & Midi Libre - \scriptsize{\underline{\texttt{midilibre.fr}}} & 183518182558 & FR \\ 
  160 & MSN France - \scriptsize{\underline{\texttt{msn.com/g00/fr-fr}}} & 136932803018290 & FR \\ 
  161 & Nice-Matin - \scriptsize{\underline{\texttt{nicematin.com}}} & 388223307574 & FR \\ 
  162 & Nord-Littoral - \scriptsize{\underline{\texttt{nordlittoral.fr}}} & 344969675415 & FR \\ 
  163 & Ouest France - \scriptsize{\underline{\texttt{ouest-france.fr}}} & 270122530294 & FR \\ 
  164 & Paris Match - \scriptsize{\underline{\texttt{parismatch.com}}} & 117714667328 & FR \\ 
  165 & Paris Normandie - \scriptsize{\underline{\texttt{paris-normandie.fr}}} & 195238257180091 & FR \\ 
  166 & R\'{e}volution Permanente - \scriptsize{\underline{\texttt{revolutionpermanente.fr}}} & 732277203520737 & FR \\ 
  167 & Sud Oest - \scriptsize{\underline{\texttt{sudouest.fr}}} & 58305334711 & FR \\ 
  168 & T\'{e}l\'{e}rama - \scriptsize{\underline{\texttt{telerama.fr}}} & 109520835773096 & FR \\ 
  169 & TF1 news - \scriptsize{\underline{\texttt{tf1.fr/news}}} & 34610502574 & FR \\ 
  170 & Var Matin - \scriptsize{\underline{\texttt{varmatin.com}}} & 365009223614 & FR \\ 
  171 & Yahoo News FR - \scriptsize{\underline{\texttt{fr.news.yahoo.com}}} & 138207559575213 & FR \\ 
  172 & Alto Adige - \scriptsize{\underline{\texttt{altoadige.gelocal.it}}} & 447795960541 & IT \\ 
  173 & Ansa - \scriptsize{\underline{\texttt{ansa.it}}} & 158259371219 & IT \\ 
  174 & Avvenire - \scriptsize{\underline{\texttt{avvenire.it}}} & 128533807252295 & IT \\ 
  175 & Corriere Adriatico - \scriptsize{\underline{\texttt{corriereadriatico.it}}} & 431943793507773 & IT \\ 
  176 & Corriere della Sera - \scriptsize{\underline{\texttt{corriere.it}}} & 284515247529 & IT \\ 
  177 & Corriere del Mezzogiorno - \newline \scriptsize{\underline{\texttt{corrieredelmezzogiorno.corriere.it}}} & 84805991975 & IT \\ 
  178 & Gazzetta di Modena - \scriptsize{\underline{\texttt{gazzettadimodena.gelocal.it}}} & 131613613524326 & IT \\ 
  179 & Gazzetta di Reggio - \scriptsize{\underline{\texttt{gazzettadireggio.gelocal.it}}} & 102328739818445 & IT \\ 
  180 & Giornale di Brescia - \scriptsize{\underline{\texttt{giornaledibrescia.it}}} & 352193836938 & IT \\ 
  181 & Giornale di Sicilia - \scriptsize{\underline{\texttt{gds.it}}} & 211307618890745 & IT \\ 
  182 & Huffington Post IT - \scriptsize{\underline{\texttt{huffingtonpost.it}}} & 276376685795308 & IT \\ 
  183 & Il Blog di Beppe Grillo - \scriptsize{\underline{\texttt{beppegrillo.it}}} & 56369076544 & IT \\ 
  184 & Il Centro - \scriptsize{\underline{\texttt{ilcentro.gelocal.it}}} & 261504285205 & IT \\ 
  185 & Il Fatto Quotidiano - \scriptsize{\underline{\texttt{ilfattoquotidiano.it}}} & 132707500076838 & IT \\ 
  186 & Il Foglio - \scriptsize{\underline{\texttt{ilfoglio.it}}} & 61703722992 & IT \\ 
  187 & Il Gazzettino - \scriptsize{\underline{\texttt{ilgazzettino.it}}} & 154142713068 & IT \\ 
  188 & Il Giornale - \scriptsize{\underline{\texttt{ilgiornale.it}}} & 323950777458 & IT \\ 
  189 & Il Giornale di Vicenza - \scriptsize{\underline{\texttt{ilgiornaledivicenza.it}}} & 154836331469 & IT \\ 
  190 & Il Manifesto - \scriptsize{\underline{\texttt{ilmanifesto.info}}} & 61480282984 & IT \\ 
  191 & Il Mattino - \scriptsize{\underline{\texttt{ilmattino.it}}} & 210639995470 & IT \\ 
  192 & Il Mattino di Padova - \scriptsize{\underline{\texttt{mattinopadova.gelocal.it}}} & 189556995002 & IT \\ 
  193 & Il Messaggero - \scriptsize{\underline{\texttt{ilmessaggero.it}}} & 124918220854917 & IT \\ 
  194 & Il Messaggero Veneto - \scriptsize{\underline{\texttt{messaggeroveneto.gelocal.it}}} & 195905383236 & IT \\ 
  195 & Il Piccolo - \scriptsize{\underline{\texttt{ilpiccolo.gelocal.it}}} & 341809745380 & IT \\ 
  196 & Il Resto del Carlino - \scriptsize{\underline{\texttt{ilrestodelcarlino.it}}} & 200174860861 & IT \\ 
  197 & Il Secolo XIX - \scriptsize{\underline{\texttt{ilsecoloxix.it}}} & 36493277214 & IT \\ 
  198 & Il Sole 24 Ore - \scriptsize{\underline{\texttt{ilsole24ore.com}}} & 38812693516 & IT \\ 
  199 & Il Tirreno - \scriptsize{\underline{\texttt{iltirreno.gelocal.it}}} & 75980429042 & IT \\ 
  200 & LA7 - \scriptsize{\underline{\texttt{la7.it}}} & 252449503661 & IT \\ 
  201 & L'Adige - \scriptsize{\underline{\texttt{ladige.it}}} & 134572506600855 & IT \\ 
  202 & La Gazzetta del Mezzogiorno - \newline \scriptsize{\underline{\texttt{lagazzettadelmezzogiorno.it}}} & 184749620911 & IT \\ 
  203 & La Gazzetta di Mantova - \scriptsize{\underline{\texttt{gazzettadimantova.gelocal.it}}} & 62769612287 & IT \\ 
  204 & La Gazzetta di Parma - \scriptsize{\underline{\texttt{gazzettadiparma.it}}} & 309928567597 & IT \\ 
  205 & La Nazione - \scriptsize{\underline{\texttt{lanazione.it}}} & 87812020989 & IT \\ 
  206 & La Nuova di Venezia e Mestre - \scriptsize{\underline{\texttt{nuovavenezia.gelocal.it}}} & 338049475695 & IT \\ 
  207 & La Nuova Sardegna - \scriptsize{\underline{\texttt{lanuovasardegna.gelocal.it}}} & 226626114877 & IT \\ 
  208 & La Provincia Pavese - \scriptsize{\underline{\texttt{laprovinciapavese.gelocal.it}}} & 57687391957 & IT \\ 
  209 & L'Arena - \scriptsize{\underline{\texttt{larena.it}}} & 108431819182401 & IT \\ 
  210 & La Repubblica - \scriptsize{\underline{\texttt{repubblica.it}}} & 179618821150 & IT \\ 
  211 & La Stampa - \scriptsize{\underline{\texttt{lastampa.it}}} & 63873785957 & IT \\ 
  212 & La Tribuna di Treviso - \scriptsize{\underline{\texttt{tribunatreviso.gelocal.it}}} & 243933437208 & IT \\ 
  213 & L'Eco di Bergamo - \scriptsize{\underline{\texttt{ecodibergamo.it}}} & 197197145813 & IT \\ 
  214 & L'Espresso - \scriptsize{\underline{\texttt{espresso.repubblica.it}}} & 259865949240 & IT \\ 
  215 & Libero Quotidiano - \scriptsize{\underline{\texttt{liberoquotidiano.it}}} & 188776981163133 & IT \\ 
  216 & L'Unione Sarda - \scriptsize{\underline{\texttt{unionesarda.it}}} & 231465552656 & IT \\ 
  217 & L'Unit\`{a} - \scriptsize{\underline{\texttt{unita.tv}}} & 292449724097 & IT \\ 
  218 & MSN Italia - \scriptsize{\underline{\texttt{msn.com/it-it}}} & 232690009759 & IT \\ 
  219 & Nuovo Quotidiano di Puglia - \scriptsize{\underline{\texttt{quotidianodipuglia.it}}} & 119992291359480 & IT \\ 
  220 & RAI News - \scriptsize{\underline{\texttt{rainews.it}}} & 124992707516031 & IT \\ 
  221 & Rai.TV - \scriptsize{\underline{\texttt{raiplay.it}}} & 88988179171 & IT \\ 
  222 & Sky TG24 - \scriptsize{\underline{\texttt{tg24.sky.it}}} & 215275341879427 & IT \\ 
  223 & TgCom24 - \scriptsize{\underline{\texttt{tgcom24.mediaset.it}}} & 40337124609 & IT \\ 
  224 & Trentino - \scriptsize{\underline{\texttt{trentinocorrierealpi.gelocal.it}}} & 82383189226 & IT \\ 
  225 & Yahoo News IT - \scriptsize{\underline{\texttt{it.notizie.yahoo.com}}} & 81262596234 & IT
\end{longtable}
\end{footnotesize}

\end{document}